# Ephemerides of Visual Binaries in Near-Parabolic Orbits

**M. I. Nouh**

**Dept. of Astronomy, National Research Institute of Astronomy and Geophysics, 11421, Helwan, Cairo, Egypt**

## Abstract

In this paper, an algorithm for the ephemerides of visual binaries in near-parabolic orbits for either elliptic or hyperbolic case has been developed. The algorithm uses the Gauss method for determining both the true anomaly and the radial distance. Applications of the algorithm are presented for the binaries ADS 11632, WDS 19553-0644, and WDS 01243-0655. The result obtained shows a good agreement and in some cases more accurate when compared with the observations as well as the predicted ephemerides.

## 1. Introduction

The study of visual binaries is almost the only way in the universe whose masses can be determined directly. Moreover, the usage of these masses leads to the discovery of the mass luminosity relationship which in turn becomes weighty support of some theories of stellar evolution.

In orbit determination of visual binaries, provisional near-parabolic orbits are used to represent the periastron section of a high-eccentricity orbit of the long and indeterminate period (Wieth-Knudsen, 1953).

What concerns us in the present paper is the computation of the ephemerides, ($\theta$ and $\rho$), where $\theta$ is the position angle (in degree) and $\rho$ is the angular separation (in seconds of arc) respectively.

In general, this type of computation plays an important role in the orbit determination of visual binaries. Because, when a set of elements is known, ($\theta$, $\rho$) at the observing times $t$ are recalculated by the ephemerides formulae, and the residuals $O$-$C$ could be found. They should be sufficiently small and mostly random distributed for an acceptable orbit.

The ephemerides formulae for these extreme cases (near parabolic orbits) are the same formulae which relate position and time in the corresponding conic section of the two bodies motion (Danby, 1988) i.e.

$$\tan\ (\theta-\Omega)=\tan\ (f\ +\omega)\cos\ i\ , \tag{1}$$

and

$$\rho=r\ \cos\ (f\ +\omega)\ \sec\ (\theta\text{-}\Omega)\ , \tag{2}$$

where $\Omega$, $\omega$, $i$, $f$ and $r$ have their usual meaning for orbits. Equations (1) and (2) convert $f$ and $r$ of the companion in the true orbit into $\theta$ and $\rho$.

The serious problem of the near-parabolic orbits for both elliptic and hyperbolic cases is due to the indeterminacy of Kepler's equation as the eccentricity $e$ tends to unity. On the other hand, as the semimajor axis $a''$ increases, both the mean anomaly and the eccentric anomaly become vanishingly small. Consequently, motion predictions of these very critical orbits can not, therefore, be treated by the conventional methods of orbit determination and need special devices (Sharaf et al., 1998).

The above mentioned importance of the computations of the ephemerides and the critical situation of near-parabolic orbits is what motivated our work: to apply the unified algorithm developed by Sharaf et al. (1998) for computing visual binaries ephemerides of near-parabolic orbits for both elliptic and hyperbolic cases. We present the applications of the algorithm to the binaries ADS 11632, WDS 19553-0644, and WDS 01243-0655.

## 2. Basic Equations

### 2.1. Gauss Method for Near Parabolic Orbits

In this section, some of the basic equations of the Gauss method and the computational developments adopted by Sharaf et al. (1998) are summarized. According to Gauss method (Battin, 1987), the true anomaly $f$ and the magnitude of the radius vector $r$ for near-parabolic orbits such that $e < 1$ are given by

$$\tan^2\frac{f}{2} = \frac{5+5e}{1+9e}\,\frac{W^2}{1-\frac{4}{5}A+C}, \tag{3}$$

and

$$r = q\,\frac{1-\frac{4}{5}A+C}{1+\frac{1}{5}A+C}\left(1+\tan^2\frac{f}{2}\right), \tag{4}$$

where $q$ is the periastron radius and is given in terms of $e$ and $a''$ as

$$q = a(1\text{-}e), \tag{5}$$

$W$ is the only one real root of the cubic equation

$$\frac{3}{B}\sqrt{\frac{\mu\,(1+9e)}{20\,q^3}}\;(t-\tau) = W^2 + 3\,W, \tag{6}$$

in which $t$, $\tau$ and $\mu$ are respectively, the time, the time of periastron passage and the gravitational constant, while $B$ is given as a function of the elliptic eccentric anomaly $E$ by

$$B = \frac{9E + \sin E}{20\sqrt{A}}, \tag{7}$$

$$A = \frac{5(1-e)}{1+9e} \quad W^2 = \frac{15(E - \sin E)}{9E + \sin E}, \tag{8}$$

Finally, $C$ is given as

$$C = \frac{A}{T} + \frac{4}{5}A - 1, \tag{10}$$

where

$$T = \tan^2 \frac{1}{2}E \quad .$$

Sharaf et al. (1998) had given $\chi^2$, $B$ and $C$ as follows

$$\chi^2 = 4A \pm \frac{8}{15}A^2 + \frac{244}{1575}A^3 \pm \frac{128}{2625}A^4 + ....., \tag{12}$$

$$B = 1 + \frac{3}{175}A^2 \pm \frac{2}{525}A^3 + \frac{471}{336875}A^4 + ...., \tag{13}$$

$$C = \frac{8}{175}A^2 \pm \frac{8}{525}A^3 + \frac{1896}{336875}A^4 + .... \tag{14}$$

## 2.2 Individual Masses and Parallax of a Binary System

The dynamical parallax ($\pi''$) and individual masses ($M_A$, $M_B$) of the components of binary systems, in solar units, can be computed (Reed, 1984) from the apparent magnitudes ($m_A$, $m_B$), the orbital period (years), and the semimajor axis of the true orbit ($a''$) in seconds of arc, via the equations

$$\log M_B = [m_B - \alpha - \frac{5}{3} \log (1 + 10^{\frac{\Delta}{\beta}}) - \lambda]/(\frac{5}{3} + \beta), \tag{15}$$

and

$$\beta = M_B (10^{\frac{\Delta}{\beta}}), \tag{16}$$

where

$$\Delta = m_A - m_B, \; \lambda = \frac{10}{3} \;\log\; P \;-\; 5 \;\log\; a'' \text{-}5, \;\; \alpha = 4.6, \;\; \beta = \text{-}9.5 .$$

Then the distance in parsecs can be determined from

$$r = \frac{(M_A + M_B)^{1/3} \; P^{2/3}}{a''} . \tag{17}$$

## 3. Computational Algorithm

In what follows, a computational algorithm for ephemerides of visual binaries of quasi-parabolic orbits for both elliptic and hyperbolic cases will be established. The algorithm is described by its purpose and its computational sequence. The algorithm is coded using Mathematica software.

- **Purpose**: To compute $(\theta, \rho)$ for a visual binary system of quasi-parabolic orbit at a given time *t*.
- **Input**: $P$, $a$, $i$, $\Omega$, $\omega$, $t$, $\tau$, $M_A$, $M_B$, Tol (specified tolerance).
- **Computational Sequence** :

1.
$$q = a(1-e) \qquad e < 1,$$
$$q = -a(1+e) \qquad e > 1.$$

2. Compute the individual masses $M_A$, $M_B$ from Equations (15) and (16).

3. Compute the parallax $\pi''$ from Equation (17).

4. Set $B=1$.

5. Solve the cubic Equation (6) for *W*.

6. Compute *A* from
$$A = \pm \frac{5(1-e)}{1+9e} W^2 .$$

7. Calculate the new value of *B* from the power series (13).

8. Repeat steps 5 to *7* until *A* ceases to change within the specified tolerance *Tol*.

9. With this value of $A$ calculate $\chi$ from power series Equation (12).

10. Calculate $C$ from the power series Equation (14).

11. Calculate $f$ from Equation (3).

12. Calculate $r$ from Equation (4)

13. Calculate $Q$ from

$$Q = \theta - \Omega = \tan^{-1}\left\{\frac{\sin\ (f+\omega)\ \cos\ i}{\cos\ (f+\omega)}\right\}.$$

14. Calculate $\theta$ from

$$\theta = Q + \Omega .$$

15. Calculate $\rho$ from

$$\rho = \frac{r\ \cos\ (f+\omega)}{\cos\ Q}.$$

16. The algorithm is completed.

## 4. Numerical Applications

The above computational algorithm is applied to obtain the ephemerides for visual binaries ADS 11632, WDS 19553-0644, and WDS 01243-0655.
The input data for the binaries taken from their respective references are listed in Table 1. The adopted constants are taken as $Tol = 10^{-15}$ and $\mu = 4\pi^2(M_A + M_B)\pi''^3$ where $\pi''$ is parallax and $M_{A,B}$ are the masses computed by the method described in Subsection 2.1.

**Table 1. Orbital elements of the selected binaries**

| Name | ADS 11632 | WDS 01243-0655 | WDS 19553-0644 |
|---|---|---|---|
| $\tau$ | 1871.53 | 1988.86 | 1974.28 |
| $P$ | 761 | 16.114 | 425 |
| $a$ | 8.11 | 0.1974 | 1.085 |
| $e$ | 1.043 | 0.927 | 0.9414 |
| $i^\circ$ | 76.74 | 116.1 | 103.05 |
| $\omega^\circ$ | 345.6 | 350.5 | 327.9 |
| $\Omega^\circ$ | 145.91 | 30.38 | 264.03 |
| $\pi''$ | 0.286 | 0.0228 | 0.0135 |
| $M_A + M_B$ | 0.696 | 2.49 | 2.85 |

| Reference | Wieth-Knudsen (1953) | Hartkopf et al. (1996) | Hartkopf et al. (1996) |
|---|---|---|---|

## 4.1. ADS 11632

The only orbit of this system has been computed by Wieth-Knudsen (1953). Table 2 shows the comparison between the *O-C's* ($\Delta\theta_{our}$, $\Delta\rho_{our}$) computed by our algorithm and those computed by using Equations (1) and (2) ($\Delta\theta_{Method\ 1}$, $\Delta\rho_{Method\ 1}$) (hereafter we shall call it Method 1).

Table 3 shows the predicted ephemerides computed by the two methods (present algorithm and Method 1). It is very clear that the comparisons are in good agreement.

**Table 2. Comparison with observations for ADS 11632.**

| $T$ | $\theta^{\circ}$ | $\rho''$ | $\Delta\theta_{our}$ | $\Delta\rho''_{our}$ | $\Delta\theta_{method\ 1}$ | $\Delta\rho''_{method\ 1}$ |
|---|---|---|---|---|---|---|
| 1905.222 | 1949.65 | 17.145 | -0.222 | 0.001 | -0.01 | 0.004 |
| 1912.53 | 151 | 17.15 | -0.398 | 0.04 | 0.1 | 0.04 |
| 1919.14 | 153.3 | 17.075 | 0.51 | 0.08 | 0.8 | 0.056 |
| 1939.616 | 156.85 | 16.414 | -0.443 | 0.095 | 0.24 | -0.1 |
| 1942.51 | 157.23 | 16.29 | -0.464 | -0.009 | -0.3 | -0.009 |
| 1948.631 | 158.86 | 16.007 | -0.56 | 0.1 | 0.1 | 0.005 |
| 1950.776 | 159.26 | 15.893 | -0.006 | -0.16 | -0.1 | -0.15 |

**Table 3. Ephemerides for ADS 11632.**

| Epoch | Present paper | | Method 1 | |
|---|---|---|---|---|
| $T$ | $\theta^{\circ}$ | $\rho''$ | $\theta^{\circ}$ | $\rho''$ |
| 1945.0 | 158.55 | 16.07 | 158.66 | 16.17 |
| 1950.0 | 159.75 | 15.83 | 159.75 | 15.94 |
| 1955.0 | 160.99 | 15.57 | 160.88 | 15.70 |
| 1960.0 | 162.28 | 15.29 | 162.05 | 15.44 |
| 1965.0 | 163.61 | 15.01 | 163.25 | 15.17 |
| 1970.0 | 164.99 | 14.72 | 164.51 | 14.90 |
| 1975.0 | 166.44 | 14.43 | 165.81 | 14.61 |
| 1980.0 | 167.94 | 14.13 | 167.17 | 14.33 |
| 1985.0 | 169.50 | 13.83 | 168.59 | 14.03 |
| 1990.0 | 171.14 | 13.52 | 171.06 | 13.74 |

### 4.2. WDS 19553-0644

This system was discovered by F. Struve in 1826. The closest separation was reached in 1974.9 (Hartkopf et al., 1996). Table 4 shows the *O-C's* computed by our algorithm and that computed by Method 1. As it is shown good agreement has been obtained. Also, the predicted ephemerides, Table 5, reveals the same result.

**Table 4. Comparison with observations for WDS 19553-0644**

| $T$ | $\theta^\circ$ | $\rho''$ | $\Delta\theta^\circ_{our}$ | $\Delta\rho''_{our}$ | $\Delta\theta^\circ_{Method\ 1}$ | $\Delta\rho''_{Method\ 1}$ |
|---|---|---|---|---|---|---|
| 1991.7152 | 107.8 | 0.296 | 1.0 | -0.002 | 1.0 | -0.002 |
| 1992.4496 | 108.5 | 0.307 | 0.29 | -0.004 | 0.29 | -0.005 |
| 1994.7080 | 106.1 | 0.348 | -0.5 | -0.005 | -0.55 | -0.006 |
| 1995.4398 | 106.22 | 0.362 | 3.07 | -0.004 | 3.08 | -0.004 |

**Table 5. Ephemerides for WDS 19553-0644**

| Epoch | Present paper | | Method 1 | |
|---|---|---|---|---|
| T | $\theta^\circ$ | $\rho''$ | $\theta^\circ$ | $\rho''$ |
| 2005 | 102.487 | 0.525 | 102.485 | 0.525 |
| 2007 | 101.967 | 0.556 | 101.966 | 0.556 |
| 2009 | 101.501 | 0.586 | 101.500 | 0.586 |
| 2011 | 101.081 | 0.615 | 101.080 | 0.615 |
| 2013 | 100.698 | 0.644 | 100.697 | 0.644 |
| 2015 | 100.347 | 0.672 | 100.346 | 0.672 |
| 2017 | 100.025 | 0.699 | 100.024 | 0.699 |

### 4.3. WDS 01243-0655

Finsen's orbit (Finsen, 1973) for this system was found to fit the portion closest to periastron better than the portion near apastron. Due to these discrepancies, Hartkopf et al. (1996) computed new orbit. This system reached a closest apparent separation in 1989.

Table 6 shows the comparison between *O-C's* computed by our algorithm and method 1. It is clearly seen that the comparison is in good agreement. Table 7 shows the predicted ephemerides computed for the epochs between 2005 and 2017. As it is shown in the table, there is a great discrepancies between our result and those computed by method 1. The behavior of the difference between two successive epochs ($\Delta\theta$, $\Delta\rho$) revealed a regular increase or decreasing for our predicted result, while there is no clear behavior for the predicted values computed by Method 1. These great discrepancies may be attributed to the high eccentricity of the orbit and low period which converts its high velocity as computed by Hartkopf et al. (1996). So we think that WDS 01243-0655 is a good example of the extreme cases which may be encountered in visual binary studies and should be treated by special methods.

**Table 6. Comparison with observations for WDS 01243-0655.**

| $T$ | $\theta^\circ$ | $\rho''$ | $\Delta\theta^\circ_{our}$ | $\Delta\rho''_{our}$ | $\Delta\theta^\circ_{Method\ 1}$ | $\Delta\rho''_{Method\ 1}$ |
|---|---|---|---|---|---|---|
| 1990.913 | 224.3 | 0.2 | 0.442 | 0.0 | 0.448 | 0.0 |
| 1990.9240 | 224.7 | 0.201 | 0.881 | 0.002 | 0.886 | 0.002 |
| 1991.7129 | 221.7 | 0.241 | 0.102 | 0.007 | 0.107 | 0.007 |
| 1991.7185 | 221.6 | 0.246 | 0.014 | 0.003 | 0.02 | 0.003 |
| 1993.9197 | 217.9 | 0.33 | 0.0 | 0.0 | 0.0 | 0.0 |
| 1994.7085 | 216.9 | 0.35 | 0.0 | 0.0 | 0.0 | 0.0 |

**Table 7. Ephemerides for WDS 01243-0655.**

| epoch | Present paper | | Method 1 | |
|---|---|---|---|---|
| $T$ | $\theta^\circ$ | $\rho''$ | $\theta^\circ$ | $\rho''$ |
| 2005 | 216.709 | 0.356 | 196.949 | 0.014 |
| 2007 | 217.832 | 0.335 | 223.948 | 0.20 |
| 2009 | 218.997 | 0.309 | 219.370 | 0.301 |
| 2011 | 220.408 | 0.246 | 203.329 | 0.167 |
| 2013 | 222.195 | 0.235 | 214.622 | 0.376 |
| 2015 | 224.386 | 0.192 | 273.491 | 0.020 |
| 2017 | 226.921 | 0.153 | 210.125 | 0.318 |

In summarizing the present paper, a unified symbolic algorithm of Gauss method is adopted for ephemerides of visual binaries of near-parabolic orbits for either elliptic or hyperbolic cases. The algorithm is applied to the two visual binaries ADS 11632 WDS 19553-0644 and WDS 01243-0655. The numerical application proved the efficiency of the developed algorithm.

## Ackenlodgment

The author is grateful to Prof. Sharaf who guiding me to the problem of near-parabolic orbits of visual binaries, to Dr. A. Saad for helpful discussions.

## References

Battin, R.H.: 1987, ***An introduction to the Mathematics and Methods of Astrodynamics***, AIAA Education Series.

Danby, J.M.A.: 1988, ***Fundamentals of Celestial Mechanics***, 2nd Edition William-Bell, Inc, Richmond.
Finsen, W., 1973, Circ. Inf. No. 61.
Hartkopf, W. I.; Mason, B. D. and McAlister, H. A., 1996, 111, 370.
Reed, B. C. : 1984, J. Roy. Astron. Soc. Can., 78, No. 2.
Sharaf, M. A.; Saad, A. S. and Sharaf, A. A., 1998, Celestial Mechanics and Dynamical Astronomy, 70, pp 201-214.
Wieth-Knudsen, N., 1953, Pub. of Lund Observatory, No. 12.